\documentclass[10pt,twocolumn,secnumarabic,amssymb, nobibnotes, aps, prd]{revtex4-2}
\usepackage{amsmath,amssymb,amsfonts}
\usepackage{graphicx}
\usepackage{bm}
\usepackage{lineno}

\newcommand{\env}[1]{\texttt{#1}}
\begin{document}
%\linenumbers

\title{Symmetry-Driven Bulk-Edge Correspondence in Electron Magnetofluids at Finite Temperature}%

\author{Xianhao Rao}%
\author{Adil Yolbarsop}\email{adil0608@ustc.edu.cn; yolbarsop@gmail.com}
\author{Hong Li}\email{honglee@ustc.edu.cn}
\author{Wandong Liu}
\affiliation{School of Nuclear Science and Technology, University of Science and Technology of China. No. 443, Huangshan Road, Hefei, Anhui, China.}
%\date{December 2018}%
\begin{abstract}
We present a theoretical framework connecting the pseudo-Chern number in momentum space to the spectral flow index in phase space for continuous media, with specific applications to topological Langmuir-cyclotron waves (TLCWs) in magnetized plasmas at uniform finite temperatures. By deriving a rigorous correspondence between these two topological invariants, we provide a solid justification for previous studies that applied this relationship heuristically across various continuous media. For magnetized plasmas with finite-temperature effects, we confirm the existence of TLCWs through numerical computation of bulk Chern number differences and analytical calculation of the spectral flow index. These findings advance the understanding of topological phenomena in continuous media.
\end{abstract}
\maketitle
%\tableofcontents

\section{Introduction}

The principles of topological insulators \cite{hasan2010colloquium, graf2013bulk, bernevig2013topological, shen2012topological}, rooted in the quantum Hall effect, have garnered significant interest in condensed matter physics due to the promise of robust mode propagation immune to backscattering, even in the presence of defects. Recently, the study of topology has expanded into continuous media, enabling applications in fluid dynamics \cite{delplace2017topological, tauber2019bulk, souslov2019topological}, acoustics \cite{xue2022topological, peri2019axial}, and plasma physics \cite{parker2020topological, fu2021topological, qin2023topological, parker2021topological, parker2020nontrivial, gao2016photonic}. While the principles of topology have proven effective across diverse physical systems, applying topological invariants to continuous systems introduces unique theoretical challenges.

In periodic lattice systems, discrete translational symmetry gives rise to integer-valued topological invariants, such as Chern numbers, defined via Berry curvature integrals over the Brillouin zone in momentum space. The bulk-edge correspondence principle ensures these invariants predict localized edge modes at interfaces between distinct topological phases. However, in continuous media, where momentum space corresponds to non-compact and contractable manifolds like \(\mathbb{R}^2\), Berry curvature integrals are often noninteger or divergent. Regardless of the numerous proposed regularizations to assign integer values to Chern-like quantities in continuous media, such bespoke integrals fail to qualify as topological invariants. The fundamental reason lies in the fact that momentum space in continuous media (\(\mathbb{R}^n\)) is topologically trivial. Consequently, these integrals cannot be regarded as true Chern numbers, as they lack the robust mathematical foundation required for a topological invariant. Consequently, phase-space topology becomes central to describing topological phenomena in such systems\cite{faure2019manifestation,jezequel2023mode,qin2023topological}.

In this context, the spectral flow index has been established by previous studies as the true topological invariant governing chiral edge modes in continuous media\cite{faure2019manifestation,jezequel2023mode,qin2023topological,fonseca2024first}. Faure’s index theorem \cite{faure2019manifestation}, an adaptation of the Atiyah-Patodi-Singer (APS) index theorem, relates the spectral flow index to the Berry monopole Chern number of eigenmode bundles over spherical surfaces that encompass Weyl points in phase space. This connection provides a rigorous framework for understanding edge modes in systems governed by phase-space topology, such as those featuring Weyl points.

Heuristic approaches employing pseudo-Chern numbers have achieved notable success in predicting topological modes. For example, in the shallow water equation model describing ocean waves, the Coriolis force induces distinct topological phases in the Northern and Southern Hemispheres, with band Chern numbers of ±1 \cite{faure2019manifestation, delplace2017topological}. The difference in the gap Chern number between the two hemispheres is 2, which corresponds to two equatorial modes: the Kelvin waves and Yanai waves. These modes exhibit a spectral flow of 2. Similarly, in the electron magnetohydrodynamics (EMHD) model, the degeneracies between Langmuir and cyclotron waves generate Berry monopoles, resulting in a gap Chern number difference of \(-1\). This difference aligns precisely with the existence of one topological Langmuir-cyclotron wave (TLCW) in spectral flow across regions with varying densities\cite{qin2023topological}.

In this study, we demonstrate that although pseudo-Chern numbers defined in momentum space are not true topological invariants, they are connected to the spectral flow index under specific conditions. By defining appropriate connections and curvatures, such as Berry connection and Berry curvature, we prove that the difference in pseudo-Chern numbers of bulk bands in momentum space equals the spectral flow index, the established true invariant. This involves regularizing eigenmodes for systems lacking symmetry in momentum space, enabling the pseudo-Chern numbers to be connected to the spectral flow index. In systems with symmetry in momentum space, this relationship holds even without regularization, despite the occurrence of noninteger bulk Chern numbers. These findings bridge the heuristic application of pseudo-Chern numbers with the rigorous framework of phase-space topology and spectral flow, offering a robust justification for previous results achieved via heuristic bulk-edge correspondence.

To investigate this correspondence, we employ the Weyl-Wigner transformation, as used in Faure’s theorem, to analyze the full-space symbol matrix of the governing PDE system. This transformation links phase-space properties to the spectral flow index via an extended clutching function. Using Stokes’ theorem, we show that the winding number of this clutching function over phase-space surfaces quantifies the spectral flow at interfaces, aligning it with the difference in pseudo-Chern numbers.

This paper is structured as follows: In Section \ref{section2}, we establish a theoretical framework to derive the spectral flow index for two energy bands using the Hermitian Wigner symbol within the Faure index theorem. The clutching function framework is extended to general phase-space surfaces, with an emphasis on the necessary conditions for connecting pseudo-Chern number differences to the spectral flow index. Section \ref{section3} applies this framework to a (2+1)-dimensional finite-temperature magnetized plasma model, highlighting how symmetry negates regularization requirements while aligning the spectral flow index with gap Chern number differences. The paper concludes with broader implications and avenues for future research in Section \ref{section4}.

\section{Establishing bulk-edge correspondence via Faure's index theorem}\label{section2}
We consider a linearized partial differential equation (PDE) governing the dynamics of a continuous medium and assume that it can be cast in a Schrödinger-like form:
\begin{equation}\label{ppde}
i\partial_t \,\xi(x,t) \;=\; \widehat{H}_\mu(x,\nabla)\,\xi(x,t),
\end{equation}
where \(\xi \in \mathbb{C}^m\) is the state vector, and \(\widehat{H}_\mu(x,\nabla): \mathbb{C}^m \to \mathbb{C}^m\) is a Hermitian operator depending on a real control parameter \(\mu\). The Wigner symbol of \(\widehat{H}_\mu(x,\nabla)\), denoted by \(H(x,k,\mu)\), is an \(m \times m\) Hermitian matrix defined over the phase space \((x,k) \in \mathbb{R}^n \times \mathbb{R}^n\). We order its eigenvalues and corresponding eigenvectors as \(\bigl(\omega_1,\xi_1\bigr), \bigl(\omega_2,\xi_2\bigr), \ldots, \bigl(\omega_m,\xi_m\bigr)\).

We assume that \(H(x,k,\mu)\) satisfies the spectral gap condition \cite{faure2019manifestation} such that for \(\|(x, k, \mu)\| \geq R_0 > 0\), the \(s\)-th and \((s+1)\)-th eigenvalues can only collapse within the region \(\|(x, k, \mu)\| \leq R_0\). By rescaling coordinates, \(R_0\) is set to 1. Outside the sphere \(B^{2n} = \{(x, k, \mu) \in \mathbb{R}^{1+2n}, \|(x, k, \mu)\| < 1\}\), a spectral gap exists in \(T = \mathbb{R}^{1+2n} \setminus B^{2n}\), bounded by constants \(M_1\) and \(M_2\) satisfying \(\omega_s(x, k, \mu) < M_1 < M_2 < \omega_{s+1}(x, k, \mu)\). The vector space spanned by the eigenvectors of the first \(s\) eigenvalues is computed using the Cauchy integral:
\begin{equation}
F = \mathrm{Ran} \oint_L \bigl(z - H(x, k, \mu)\bigr)^{-1} dz,
\end{equation}
where \(L\) is a contour in the complex plane enclosing \(\omega_1, \omega_2, \ldots, \omega_s\). The vector space \(F\) is interpreted as an \(s\)-dimensional complex vector bundle over \(T\).

We define the sets \(P^0\), \(P^+\), and \(P^-\) as follows:

\[
P^0 = \{(x,k,\mu) \in \mathbb{R}^{1+2n} \mid \mu = 0\},
\]
\[
P^+ = \{(x,k,\mu) \in \mathbb{R}^{1+2n} \mid \mu > 0\},
\]
\[
P^- = \{(x,k,\mu) \in \mathbb{R}^{1+2n} \mid \mu < 0\}.
\]
Let \(\Gamma\) be an arbitrary closed surface that contains \(B^{2n}\) internally and is homotopic to the boundary of \(B^{2n}\), denoted as \(S^{2n}\).  The thick-shell region between \(\Gamma\) and \(S^{2n}\) is denoted as \(M\). A generalized clutching function in \(U(r)\) for the bundle \(F\), denoted by \(U(x,k)\), is constructed over \(A^0\equiv M\cap P^0\). This function glues the local trivialization of \(F\) defined on \(A^+\equiv M\cap P^+\) with that on \(A^-\equiv M\cap P^-\).
Let \(\{\xi_1,\xi_2,\ldots,\xi_r\}\) represent the local trivialization of \(F\) in the closure of \(A^+\), and \(\{\xi_1',\xi_2',\ldots,\xi_r'\}\) represent the trivialization in the closure of \(A^-\). At the equator \(A^0\), the relationship
\[
(U(x,k))_{ij}\xi_j' = \xi_i
\]
holds by definition of \(U(x,k)\). Under the induced topology in \(\mathbb{R}^{2n+1}\), the intersections \(\Gamma\cap P^\pm\) and \(S^{2n}\cap P^\pm\) are trivial. Therefore, \(U(x,k)|_{\Gamma\cap P^0}\) and \(U(x,k)|_{S^{2n}\cap P^0}\) act as the clutching functions over \(\Gamma\) and \(S^{2n}\), respectively.

Extending the clutching function \(U(x,k)|_{S^{2n} \cap P^0}\) to the entire phase space \(\mathbb{R}^n \times \mathbb{R}^n\) as follows:
\begin{widetext}
\begin{equation}\label{U-tilde}
\widetilde{U}(x,k) =
\begin{cases}
    U(x,k), & \text{if } (x,k) \in S^{2n} \cap P^0, \\ \\
    \|(x,k)\|U\left(\frac{(x,k)}{\|(x,k)\|} \right), & \text{if } (x,k) \in P^0 - S^{2n},
\end{cases}
\end{equation}
\end{widetext}
the Weyl quantization of \(\widetilde{U}(x,k)\), denoted by \(\widehat{\tilde{U}} \equiv Op_\epsilon(\tilde{U})\), is an elliptic operator. Its algebraic index is expressed as:
\begin{equation}
\mathrm{Ind} \widehat{\tilde{U}} = \dim \mathrm{Ker} \widehat{\tilde{U}} - \dim \mathrm{Ker} \widehat{\tilde{U}}^*.
\end{equation}

According to Faure's theorem \cite{faure2019manifestation}, or equivalently, the Berry-Chern monopole and spectral flow correspondence \cite{delplace2022berry}, the spectral flow index \(N\), quantifying the net number of upward crossings of eigenmodes between bands \(\omega_s\) and \(\omega_{s+1}\) as the control parameter \(\mu\) traverses the real line, is precisely equal to \(\mathrm{Ind} \widehat{\tilde{U}}\), which in turn coincides with the Chern number \(C\) defined on the sphere \(S^{2n}\):
\begin{equation}\label{Chern-1}
C = \int_{S^{2n}} \frac{1}{n!} \left(\frac{i}{2\pi}\right)^n \mathrm{Tr} R^{\wedge n},
\end{equation} 
where \(R\) denotes the curvature tensor, a matrix-valued two-form associated with the vector bundle \(F\). In the specific case of \(n=1\), this curvature simplifies to the well-known Berry curvature associated with the eigenstate \(\xi_i\) for each band \(\omega_i\) \cite{simon1983holonomy,faure2019manifestation}. This fundamental relationship can be succinctly summarized as:
\begin{equation}\label{FIT}
N = C.
\end{equation}

Since \(F\) is a rank-\(s\) vector bundle on \(S^{2n}\), an alternative form of Chern number, \(C\), in term of winding number \cite{faure2019manifestation,prodan2016bulk} is:
\begin{equation}
C = W_{2n-1} = c_n \int_{S^{2n-1}} \mathrm{Tr} (\tilde{U}^{-1} d\tilde{U})^{2n-1},
\end{equation}
where \(c_n = -2n!/((2n)!(2\pi i)^n)\) and \(S^{2n-1} = S^{2n} \cap P^0\). As per Eq.(\ref{U-tilde}), recognizing that \(U = \tilde{U}\) on \(S^{2n-1}\), we obtain:
\begin{equation}\label{Chern-3}
C = W_{2n-1} = c_n \int_{S^{2n-1}} \mathrm{Tr} (U^{-1} dU)^{2n-1}.
\end{equation}

It follows from applying Stokes' theorem to \(M\) that :
\begin{equation}\label{stokes}
\begin{aligned}
0 & = \int_M \mathrm{dTr}(U^{-1} dU)^{2n-1} \\
&= c_n \int_{\Gamma \cap P^0} \mathrm{Tr}(U^{-1} dU)^{2n-1} \\
&- c_n \int_{S^{2n-1}} \mathrm{Tr}(U^{-1} dU)^{2n-1}.
\end{aligned}
\end{equation}
The first equality in equation (\ref{stokes}) arises because \(\mathrm{Tr}(U^{-1} dU)^{2n-1}\) is a closed form due to the anti-symmetry of one-form commutations (see Appendix \ref{Closed_form}). The second term on the r.h.s of the second equality is precisely the Chern number in Eq.(\ref{Chern-3}).

Hence, the spectral flow index \(N\) can be expressed, following from Eq.(\ref{FIT}), (\ref{Chern-3}) and (\ref{stokes}), in terms of a winding number: 
\begin{equation}
    N = c_n \int_{\Gamma \cap P^0} \text{Tr}(U^{-1} dU)^{2n-1}.
\end{equation}
Since the integral on the r.h.s. of the above equation characterizes the topology of the nontrivial vector bundle \(F\) over \(\Gamma\) as well, it thereby follows:
\begin{equation}\label{SFI-2}
    N = \int_{\Gamma} \frac{1}{n!} \left(\frac{i}{2\pi}\right)^n \text{Tr} R^{\wedge n}.
\end{equation}
This result enables us to apply appropriate homotopic deformation to the sphere, $S^{2n}$, to derive the desired bulk-edge correspondence for a continuous media system. It is also worth mentioning that, combining results from Eq.(\ref{Chern-1}), (\ref{FIT}) and (\ref{Chern-3}), it is easy to derive a generalized version of the boundary isomorphism theorem, which states that the Chern number of the eigenbundle F remains the same on two homotopic surfaces. Specifically, when \(\Gamma\) is a sphere centered at the origin enclosing all degeneracies, this result aligns with the boundary isomorphism theorem introduced in Ref.\cite{qin2023topological}.

Up to this point, the discussion has remained general. By applying homotopic deformations to the sphere \( S^{2n} \), various correspondences between topological quantities can be derived. To apply above established results to the ``warm" magnetized plasma system discussed in the next section, it is advantageous to focus on a (2+1)-dimensional system and explicitly formulate these correspondences.

\begin{figure}
    \centering
    \includegraphics[width=0.5\textwidth]{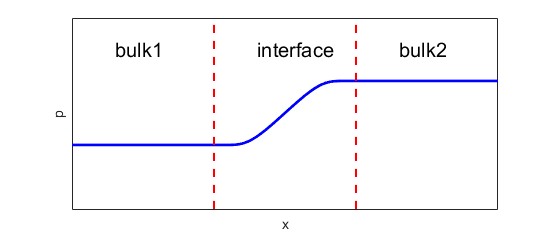}
    \caption{Schematic representation of the bulk interface in a continuous medium. The \(x\)-axis denotes spatial position, while the \(p\)-axis represents the parameter values on which the operator \(\widehat{H}(x,\nabla)\) depends. In each bulk region, \(\widehat{H}\) remains uniform.}
    \label{fig1}
\end{figure}

Consider the reduced (2+1)-dimensional problem shown in Figure \ref{fig1}. Here, two uniform bulks with distinct uniform parameters are interfaced, and the parameter varies continuously along the interface's spatial coordinate \(x\). Applying a Fourier transform along the uniform \(y\)-direction introduces the wave number \(k_y\), reducing the system operator to \(\widehat{H}(x,\partial_{x},k_{y})\).

We assume that \(\widehat{H}(x,\partial_{x},k_{y})\) is locally spin-\(1/2\), implying that its symbol near degenerate points can be approximated as a \(2 \times 2\) Hermitian matrix. In the bulk regions, where the parameters of \(\widehat{H}(x,\partial_{x},k_{y})\) are uniform and independent of \(x\), the symbol \(H(x,k_{x},k_{y})\) satisfies \(\partial_x H(x,k_{x},k_{y}) = 0\), as illustrated in Figure \ref{fig1}.

The parameter \( k_y \) acts as \( \mu \) in the symbol \( H(x,k,\mu) \). Suppose the interface between the two bulks is confined to the region \( x \in [-d, d] \). We assume that all degeneracies are located within this interface and that, sufficiently far from the origin, \( H(x,k_x,k_y) \) maintains a band gap larger than a constant.

\begin{figure}
    \centering
    \includegraphics[width=1\linewidth]{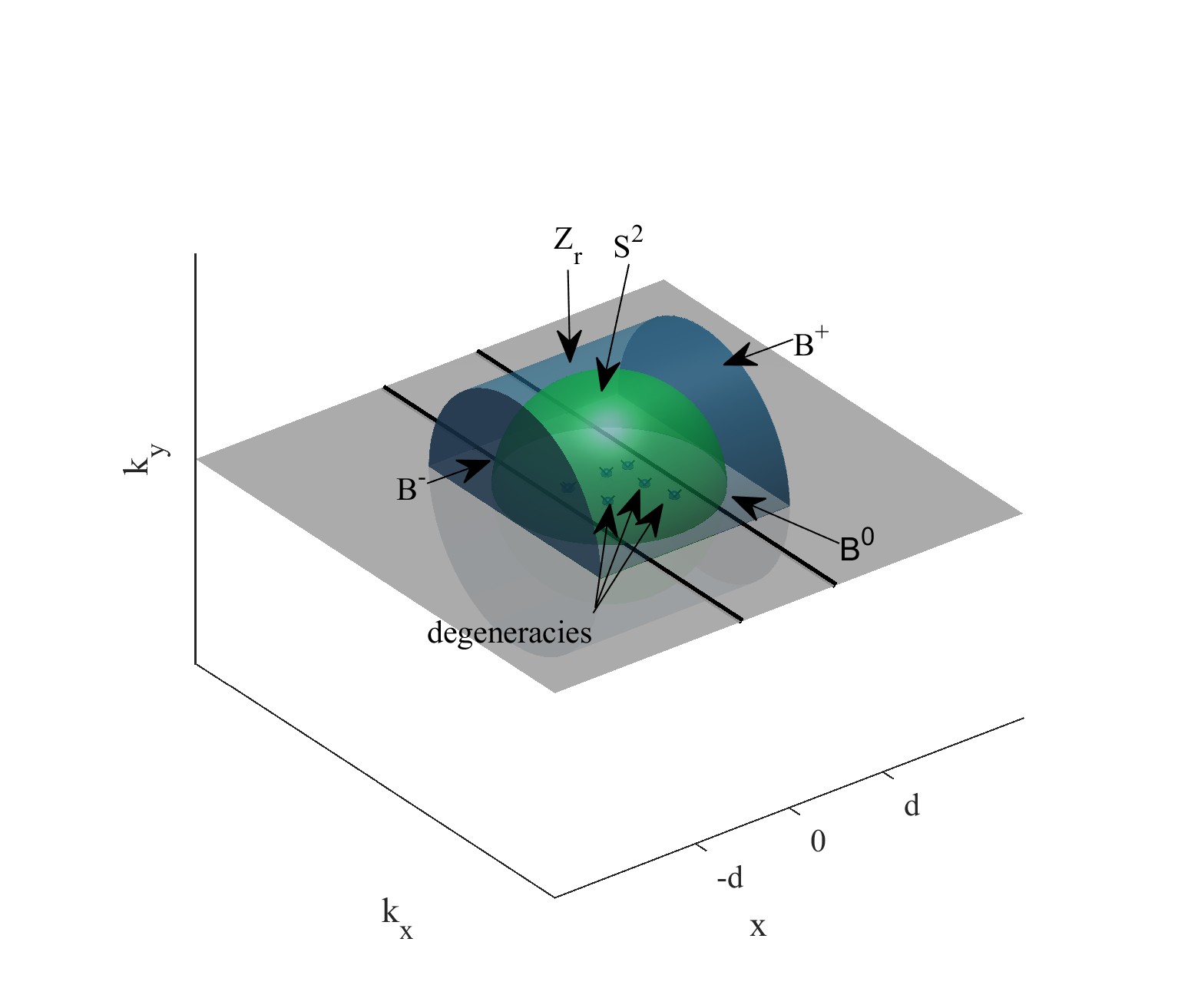}
    \caption{Depiction of \(Z_r\) in phase space. The blue cylinder represents \(Z_r\), whose circular face at \(x=d\) is labeled \(B^+\) and at \(x=-d\) is labeled \(B^-\), with the curved side boundary denoted by \(B^0\). Inside the cylinder lies the green sphere \(S^2\), encompassing all degeneracies of the symbol \(H(x,k_x,k_y)\).} 
    \label{fig:enter-label}
\end{figure}

We specify the deformed closed surface \(\Gamma\) as the cylinder \( Z_r \), ensuring it contains all degeneracies inside. The cylindrical region \( Z_r \) is defined as:
\[
Z_r = \partial\{(x,k_x,k_y), -d < x < d\} \cap \{(x,k_x,k_y), k_x^2 + k_y^2 \leq r^2\}.
\]
The boundary sections of the cylinder can be written as:
\[
B^+ = \{(x,k_x,k_y) \mid k_x^2 + k_y^2 \leq r^2 \text{ and } x = d\},
\]
\[
B^- = \{(x,k_x,k_y) \mid k_x^2 + k_y^2 \leq r^2 \text{ and } x = -d\},
\]
\[
B^0 = \{(x,k_x,k_y) \mid k_x^2 + k_y^2 = r^2 \text{ and } -d < x < d\}.
\]
Thus, the region \( Z_r \) is expressed as:
\[
Z_r = B^+ \cup B^- \cup B^0,
\]
as shown in Fig.(\ref{fig:enter-label}).

Since we are considering a spin-\(1/2\) model with \(n=1\), the curvature 2-form on the right-hand side of the equation (\ref{SFI-2}) can be computed using the Berry curvature. Specifically, for the eigenstate \(\xi\) corresponding to the lower band of the symbol \(H(x,k_x,k_y)\), the Berry curvature \(F\) is given by:
\begin{equation}
    F = \frac{i}{2\pi} \text{Tr} R = \langle \mathrm{d}\xi \wedge | \mathrm{d}\xi \rangle.
\end{equation}
The Berry curvature is summed across each band if multiple non-degenerate lower bands are present.
The integral in equation (\ref{SFI-2}) over \(Z_r\) can be divided into three parts:
\begin{equation}
    N = \frac{i}{2\pi} \int_{B^+} \mathrm{Tr} R + \frac{i}{2\pi} \int_{B^-} \mathrm{Tr} R + \frac{i}{2\pi} \int_{B^0} \mathrm{Tr} R.
\end{equation}
In the limit where the radius of the cylinder \(Z_r\) approaches infinity,  \(r \to \infty\), and if the integral over \(B^0\) becomes negligible, the index reduces to
\begin{equation}\label{SFI-4}
    N \;=\;\frac{i}{2\pi}\int_{B^+}\mathrm{Tr}\,R \;+\;\frac{i}{2\pi}\int_{B^-}\mathrm{Tr}\,R.
\end{equation}
By consistently orienting \(B^+\) and \(B^-\) in terms of \((k_x, k_y)\), the relation presented in (\ref{SFI-4}) can be expressed in a more familiar form: the bulk-edge correspondence:
\begin{equation}\label{BEC}
 N \;=\;C^+ \;-\; C^-,   
\end{equation}
where \(C^+\) and \(C^-\) denote the bulk Chern numbers computed over the momentum space \((k_x,k_y)\) on the respective sides \(B^+\) and \(B^-\). We continue to refer to these as “bulk Chern numbers,” even though—strictly speaking—they are not the true Chern numbers and thus are not topological invariants in the rigorous sense, as Chern numbers are not defined in the momentum space of continuous media. Nevertheless, for simplicity, we retain the term “bulk Chern numbers.” Both \(C^+\) and \(C^-\) are obtained by integrating the Berry curvature of the energy bands below the degenerate points.

The result presented in Eq.~(\ref{BEC}) embodies the bulk-edge correspondence, linking the genuine topological invariant in phase space—namely, the spectral flow index—to the bulk Chern numbers (or pseudo-Chern numbers) evaluated in two separate bulk regions in momentum space. Two key conditions are required for this correspondence. First, the Berry curvature must be used to compute the bulk Chern numbers. Second, the contribution of the curvature integral over \(B^0\) must vanish as the radius of the cylinder goes to infinity. In continuous media, this second condition is typically satisfied through various methods of Berry curvature regularization, ensuring that the integral on \(B^0\) is zero\cite{delplace2017topological,fu2021topological,parker2020nontrivial}. However, in systems featuring certain symmetries \cite{parker2020nontrivial,fu2021topological}, the integral over \(B^0\) can vanish on its own, which means Eq.~(\ref{BEC}) continues to hold even without explicitly regularizing the curvature. The “warm” plasma system discussed in the following section provides another example exhibiting this type of symmetry.

When a degenerate point exists at \(k = \pm \infty\), the integral on the right-hand side of the equation may diverge from the spectral flow index derived from \(\widehat{H}\). This difference is attributed to an extra contribution to the spectral flow index at infinity, known as a ghost mode \cite{graf2021topology}. Additionally, violations of bulk-boundary correspondence can occur in systems with sharp boundaries \cite{bal2024topological}, where the discontinuity in \(\widehat{H}\) disrupts the semi-classical framework. In this case, even with integrable \(C^+\) and \(C^-\), contributions from \(B^0\) become significant and non-negligible.

As stated previously, the bulk-edge correspondence expressed in Eq.(\ref{BEC}) holds, provided that the total integral over \(B^0\) approaches zero as \(|k| \to \infty\). In one case, the cancellation occurs when \(H(x,k_x,k_y)\) shows symmetries that cause the Berry curvature contributions on \(B^0\) to vanish. For instance, in the linearized EMHD framework, where a constant matrix commutes with its Wigner symbol, Berry curvatures at opposite \(k\)-values cancel due to symmetry. This cancellation is demonstrated in Appendix \ref{Berry}. In the next section, we will examine the TLCW in magnetized plasma with finite temperature, or warm magnetized plasma, using the correspondence relation obtained in this section and a two-level approximation.

\section{Berry curvature integral of isothermal electron magnetofluids and spectral flow’s calculation}\label{section3}

We extend the magnetized cold plasma model by incorporating finite temperature effects to analyze the TLCWs, building upon previous studies of these waves in magnetized cold plasma \cite{fu2021topological,parker2021topological,qin2023topological}. The dynamics of warm electron fluids can be described by the following system of equations \cite{zhao2017properties}:
\begin{equation}\begin{gathered}
m_en\partial_t\boldsymbol{v}=-ne(\boldsymbol{E}+\boldsymbol{v}\times \boldsymbol{B})-\nabla P_e, \\
\partial_tn=-n\nabla\cdot \boldsymbol{v}-\nabla n\cdot \boldsymbol{v}, \\
\partial_{t}\boldsymbol{B}=-\nabla\times \boldsymbol{E}, \\
\partial_t\boldsymbol{E}=c^2\nabla\times \boldsymbol{B}+c^2\mu_0ne\boldsymbol{v} .
\end{gathered}\end{equation}
Here, \(\boldsymbol{v}\) represents the electron flow velocity, \(n\) the electron density, \(\boldsymbol{E}\) the electric field, and \(\boldsymbol{B}\) the magnetic field. The electron pressure \(P_e\) follows the isothermal relation \(P_e=nk_BT_e\), where \(k_B\) is the Boltzmann constant and \(T_e\) is the constant electron temperature.

The equilibrium configuration satisfies \(en\boldsymbol{E}_0 + \nabla P_{e_{0}} = 0\), with a uniform magnetic field \(\boldsymbol{B}_0 = B_z\boldsymbol{e}_z\) and zero equilibrium flow (\(\boldsymbol{v}_0 = 0\)). We restrict the inhomogeneity of equilibrium quantities to the \(x\) direction, such that \(n_0 = n_0(x)\) and \(\boldsymbol{E}_0 = E_0(x)\boldsymbol{e}_x\). 

Following standard linearization and normalization procedures (detailed in Appendix \ref{linearization}), we obtain a Schrödinger-like equation:
\begin{equation}\label{sch}
    i\partial_t \Psi = \widehat{H} \Psi.
\end{equation}
In our normalization scheme, time is normalized to the cyclotron frequency \(\sigma = {eB}/{m_e}\), spatial variables to \({c}/{\sigma}\), electric field to a reference field \(E\), magnetic field to \(E/c\), velocity to \(\sqrt{\epsilon_0 E^2/(m_e n_0)}\), and density to \(\sqrt{n_0 \epsilon_0 E^2/(k_B T_e)}\). The state vector \(\Psi\) is defined as:

\begin{equation}
\Psi = \begin{pmatrix} \tilde{v}_1 \\ \tilde{E}_1 \\ \tilde{B}_1 \\ \tilde{n}_1 \end{pmatrix},
\end{equation}
which is a \(10\times1\) vector. The Hamiltonian operator takes the form:
\begin{equation}\label{H_linear}
\widehat{H} = \begin{pmatrix} 
i\sigma \boldsymbol{e}_\mathbf{z} \times & -i\omega_p \boldsymbol{I}_3 & 0 & \frac{1}{2}\frac{iu}{L_{n_0}} - ui\nabla \\ 
i\omega_p \boldsymbol{I}_3 & 0 & i\nabla \times & 0 \\ 
0 & -i\nabla \times & 0 & 0 \\ 
-\frac{1}{2}\frac{iu}{L_{n_0}} - ui\nabla & 0 & 0 & 0 
\end{pmatrix},
\end{equation}
where \(\omega_p = \sqrt{(n_0 e^2/(m_e \epsilon_0))}/(eB /m_e)\) is the normalized electron plasma frequency and \(u = \sqrt{(k_B T_e/m_e)}/c\) is the normalized electron thermal velocity. The electron density scale length function \(L_{n_0}(x)\) is determined by the equilibrium and isothermal conditions:
\begin{equation}
L_{n_0}^{-1}(x) = \frac{\nabla n_0}{n_0} = \frac{\nabla P_{e_0}}{k_B T_e n_0} = -\frac{eE_0}{k_B T_e}.
\end{equation}

To analyze the spectrum flow, we leverage the fact that \(L_{n}^{-1}\) varies only in the \(x\) direction due to the inhomogeneity of \(n_0\). It is convenient to apply a Fourier transform in the \(y\) and \(z\) spatial coordinates and time \(t\):
\begin{equation}
\Psi = \psi(x,k_y,k_z)\exp[i(k_yy + k_zz - \omega t)].
\end{equation}
The equation (\ref{sch}) reduces to:
\begin{equation}
    i\partial_t \psi = \widehat{H}(x,\partial_x,k_y,k_z) \psi.
\end{equation}

For our analysis, we treat \(k_z\) as a fixed constant and take \(u < 1\), which is a reasonable assumption since the thermal velocity of the plasma typically remains well below the speed of light. Applying the Weyl-Wigner transform in the \(x\) direction yields the Wigner symbol \(H(x,\boldsymbol{k})\) of \(\widehat{H}(x,\partial_x,k_y,k_z)\):
\begin{equation}\label{H0handH1}
   H(x,\boldsymbol{k})=H_0 (x,\boldsymbol{k})+H_1 (x,\boldsymbol{k}), 
\end{equation}
where \(H_0(x,\boldsymbol{k})\) is defined as:
\begin{equation}
H_0(x,\boldsymbol{k})=\begin{pmatrix}i\boldsymbol{e}_z\times&-i\omega_p\boldsymbol{I}_3&0&u\boldsymbol{k}^T\\i\omega_p\boldsymbol{I}_3&0&-\boldsymbol{k}\times&0\\0&\boldsymbol{k}\times&0&0\\u\boldsymbol{k}&0&0&0\end{pmatrix},
\end{equation}
and \(H_1(x,\boldsymbol{k})\) is given by:
\begin{equation}
H_1(x,\boldsymbol{k})=\begin{pmatrix}0&0&0&\frac{iu}{2L_{n_0}}\\0&0&0&0\\0&0&0&0\\-\frac{iu}{2L_{n_0}}&0&0&0\end{pmatrix}.
\end{equation}

The Wigner symbol \(H(x,\boldsymbol{k})\) preserves its Hermitian nature, consistent with the magnetized cold plasma model, where its upper-left \(3 \times 3\) block defines the system’s dynamics. In the limit \(u \to 0\), the system simplifies to the magnetized cold plasma case discussed in \cite{fu2021topological,qin2023topological,parker2021topological}.

We define a new interpolated symbol:
\begin{equation}\label{interpo-H}
    H_t(x,\boldsymbol{k}) = H_0(x,\boldsymbol{k}) + t H_1(x,\boldsymbol{k}), \quad t \in [0,1].
\end{equation}
With finite-temperature effects included, \(H_t(x,\boldsymbol{k})\) retains particle-hole symmetry \(H^*(x,-\boldsymbol{k}) = -H(x,\boldsymbol{k})\) \cite{parker2021topological,jin2016topological,fu2021topological}. Its dispersion relation is invariant under \(\boldsymbol{k} \to -\boldsymbol{k}\), ensuring symmetry in the energy bands with respect to \(\omega = 0\). Thus, if \(\omega\) is an eigenvalue, so is \(-\omega\).

As detailed in Appendix (\ref{Berry}), the nonzero eigenvalues of \(H_t(x,\boldsymbol{k})\) are denoted as \(\pm\omega_1, \pm\omega_2, \pm\omega_3, \pm\omega_4\), ordered such that \(0 < \omega_1 < \omega_2 < \omega_3 < \omega_4\). If \(L^{-1}(x)\) is bounded, as \(k_x\) or \(k_y \to \infty\), the eigenvalues asymptotically approach:
\[
(\omega_1, \omega_2, \omega_3, \omega_4) \to (k_\bot^s, uk_\bot, k_\bot, k_\bot),
\]
where \(k_\bot = \sqrt{k_x^2 + k_y^2}\) and \(s < 0\).
Thus, the gap condition between \(\omega_1\) and \(\omega_2\) persists throughout the interpolation range \(t \in [0,1]\), ensuring stability in the spectral gap.

Assuming that the density $n_0(x)$ follows the distribution outlined in Figure \ref{fig1}, we find that within the bulk regions, where the density is approximately constant, the quantity \(L_{n_0}^{-1}\) equals zero. Consequently, \(H_1(x, \boldsymbol{k})\) vanishes in these regions, and thus \(H_t\) simplifies to \(H_t = H_0\) for all $t \in [0,1]$. 
The total Chern number associated with bands below the gap between \( \omega_1\) and \(\omega_2\) is expressed as:
\begin{equation}\label{Chern_H}
 N = C^+ - C^- + \frac{i}{2\pi} \int_{B^0} \text{Tr}R.
\end{equation}

Employing the Berry curvature to compute \(N\) in Eq.~(\ref{Chern_H}), note that the first two terms in Eq.~(\ref{Chern_H}), corresponding to the bulk Chern numbers \(C^+\) and \(C^-\), are determined entirely by \(H_0\). This is because they involve integrals evaluated over uniform bulk regions in phase space, where \(H_1 = 0\). For the last term in Eq.~(\ref{Chern_H}), since \(H_0\) satisfies the symmetry \(PH_0(x, -k_x, -k_y) = H_0(x, k_x, k_y)P\), where \(P\) is a constant diagonal matrix, the contribution of \(H_0\) to the integral over \(B^0\) vanishes (see Appendix~\ref{Berry}). 

Now consider the full interpolated symbol \(H_t\), where \(t\) varies from 0 to 1. This variation establishes a homotopy between the eigenbundles of \(H_0\) and \(H_0 + H_1\) formed by the eigenvectors below the gap between \(\omega_1\) and \(\omega_2\). Importantly, no additional degeneracy arises between any two nonzero bands of the symbol \(H_t\) for sufficiently large \(\|(x, k_x, k_y)\|\). As a result, the spectral flow index \(N\) remains an invariant integer as \(t\) varies from 0 to 1. Combining this with the condition that \(H_0 + H_1\) reduces to \(H_0\) in the bulk regions, the Chern number of \(H_0 + H_1\) remains \(C^+ - C^-\). Therefore, the overall spectral flow index of the system is preserved.

\begin{figure}
    \centering
    \includegraphics[width=0.5\textwidth]{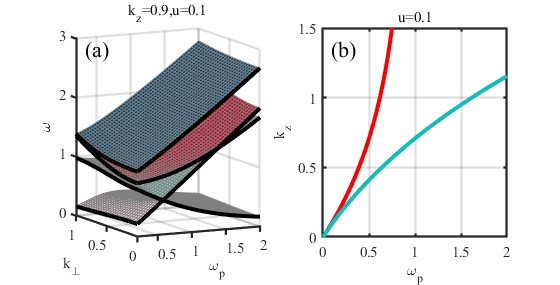}
    \caption{Dispersion relations of \(H_0\).  
(a) Positive energy bands of \(H_0\) in phase space for \(k_z = 0.9\) and \(u = 0.1\).  
(b) The degeneracies between the Langmuir wave and cyclotron waves in the \(k_z\text{-}\omega_p\) plane.}
    \label{fig2_1}
\end{figure}
%Hereafter, we consider the eigenbundle is generated by the bands below the Fermi surface in the gap of Langmuir wave and electron cyclotron wave. 
To determine the band spectral flow index of \(H\), we calculate \(C^\pm\) using fixed parameters \(k_z=0.9\) and \(u=0.1\). 

Figure \ref{fig2_1}(a) shows the energy band distribution of \(H_0\)'s symbol as a function of \(k\) and \(\omega_p\). At the cross-section where \(k_\perp=0\), we observe two degenerate points (Weyl points) in the energy bands. These points represent two distinct resonances: one between the R-wave in cyclotron waves and the Langmuir wave (intersection of pink and light green bands), and another between the Langmuir wave and the L-wave in cyclotron waves (intersection of red and light green bands).

The calculation of \(C^\pm\) requires integrating the Kubo formula over the \(k_x-k_y\) plane in two bulk regions:
\begin{equation}
    C^+-C^-=\lim_{r\to+\infty}(C_{E_F}^+(r)-C_{E_F}^-(r)):=\lim_{r\to+\infty}N(r).
\end{equation}
The gap Chern number in the bulk regions, representing the total Berry curvature integral of all eigenvectors of \(H_0(x, k)\) below the Fermi surface, is given by:
\begin{widetext}
\begin{equation}
C_{E_F}^\pm(r)=\frac{-1}{2\pi i}\sum_{\omega<E_F,\omega^{\prime}>E_F}\int_{k_x^2+k_y^2<r^2}\left(\frac{(\Psi_\omega^*\nabla_kH\Psi_{\omega^{\prime}})\times(\Psi_{\omega^{\prime}}^*\nabla_kH\Psi_\omega)}{(\omega-\omega^{\prime})^2}\right)_{\omega_p=\omega_p^\pm}.
\end{equation}
\end{widetext}

For the analysis of R-wave resonance Weyl points shown in Figure \ref{fig2_1}(a), we perform numerical integration of \(C_{E_F}^\pm\) using \(\omega_p^-=0.1\) in bulk1 and \(\omega_p^+=0.9\) in bulk2. The plasma frequency profile across the interface is taken to be:
\begin{equation}
    \omega_p(x)=\frac{\omega_p^-+\omega_p^+}{2}+\frac{\omega_p^+-\omega_p^-}{2}\tanh{\frac{x}{\delta}},
\end{equation}
where the interface region spans \(x\in[-10, 10]\), and \(\delta=1\) is used for calculating the Berry curvature integral over \(B^0\). The Fermi surface for this analysis is chosen to lie between the pink and light green bands.
\begin{figure}
    \centering
    \includegraphics[width=1\linewidth]{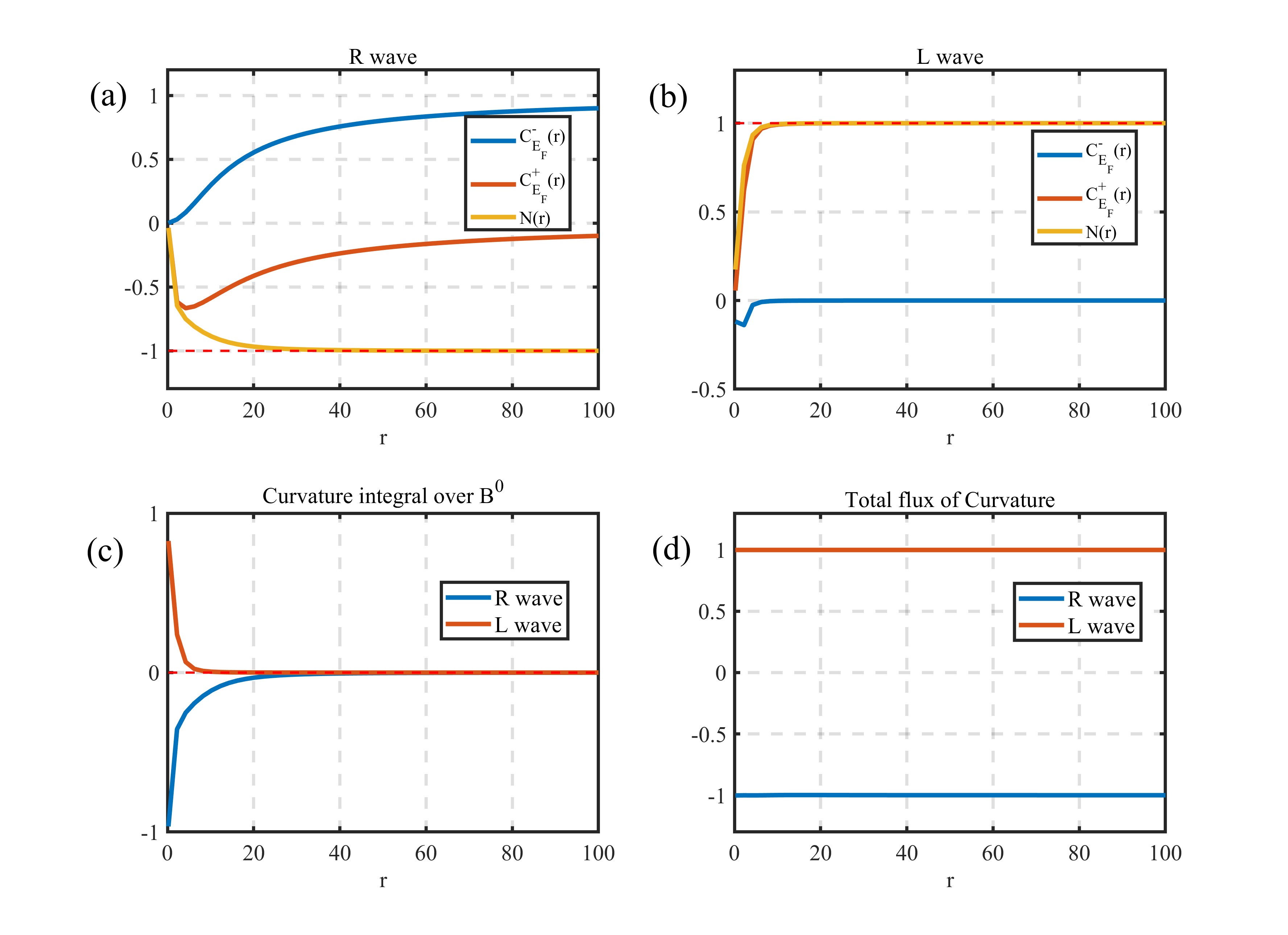}
    \caption{Chern number as a Berry curvature integral.
(a) The bulk Chern numbers of the R wave, and their difference, for \(\omega_p^- = 0.1\) in bulk 1 and \(\omega_p^+ = 0.9\) in bulk 2.  
(b) As in (a), but for the L wave with \(\omega_p^- = 1\) and \(\omega_p^+ = 2\).  
(c) Berry curvature flux through \(B^0\) using the same parameters as in (a) and (b).  
(d) The total Berry curvature flux over \(B^\pm\) and \(B^0\).}
    \label{fig:3}
\end{figure}

Figure \ref{fig:3} illustrates the convergence of the total Berry curvature integral in the region \( B^\pm \) of bands located below the Fermi surface to the gap Chern number as the integral region's radius \( r \) approaches infinity. The blue curve represents the Chern number for bulk 1 as \( r \) tends to infinity, while the orange curve corresponds to bulk 2. The yellow curve depicts the difference in Berry curvature between the two bulk bands. In the bulk, the difference in gap Chern number approaches an integer limit of $-1$ as the integral radius \( r \) increases, indicating that the spectral flow index is $-1$ and thus there is one net downward eigenmodes, TLCW, in spectral flow in the gap.
\begin{figure}
    \centering
    \includegraphics[width=0.4\textwidth]{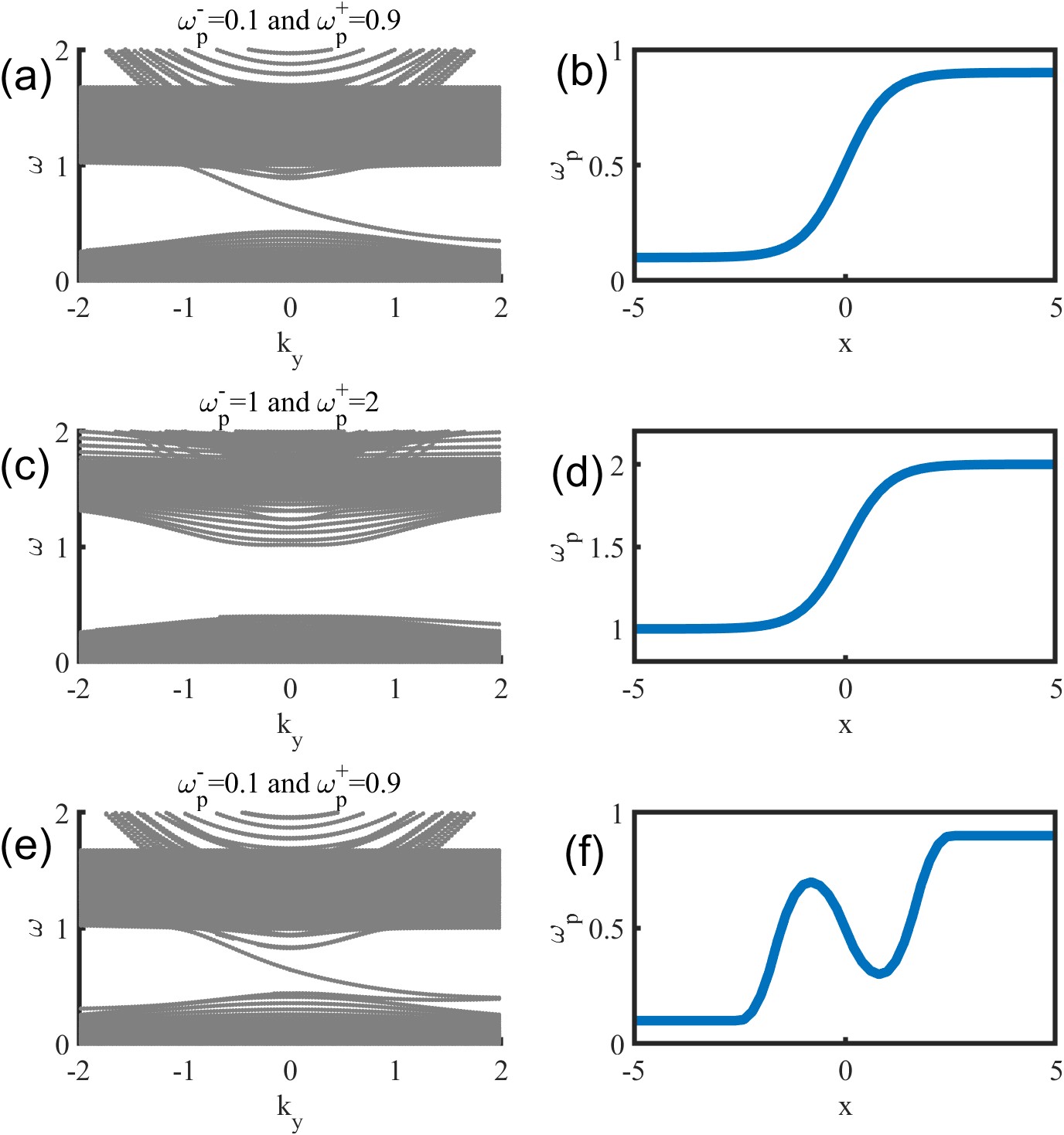}
    \caption{Band Structure for Different Plasma Density Profiles at $k_z=0.9$ and $u=0.1$. (a) Band structure $\omega(k_y)$ corresponding to the density profile shown in (b), where the bulk density settings are $\omega_p^-=0.1$ and $\omega_p^+=0.9$. (c) Band structure $\omega(k_y)$ associated with the density profile in (d), where the bulk density settings are $\omega_p^-=1$ and $\omega_p^+=2$. (e) Band structure $\omega(k_y)$ linked to the density profile in (f), where the bulk density settings are again $\omega_p^-=0.1$ and $\omega_p^+=0.9$.}
    \label{fig4}
\end{figure}
Figures \ref{fig4}(a-b) demonstrate that at the interface of the two bulk structures, with \( \omega_p^-=0.1 \) and \( \omega_p^+=0.9 \), there are chiral modes present at fixed \( k_y \) with a negative group velocity (\( d\omega/dk_y < 0 \)), which directly corresponds to the spectral flow index of -1. 

It is also noticeable that although neither of the two gap Chern numbers \( C^\pm \) is an integer, their difference is. Furthermore, if we consider \( \omega_p^-=1 \) for bulk 1 and \( \omega_p^+=2 \) for bulk 2, we find that the spectral flow index of L-wave resonance Weyl points is +1, as shown in Figure \ref{fig:3}(b). These results are consistent with the analytic treatment via two-band approximation, which is employed to calculate the curvature flux through a vanishing sphere surrounding an isolated Weyl point (see Appendix \ref{two_band}). 

For magnetized plasma, when the density (and thus the plasma frequency) varies monotonically at the interface, the Langmuir wave of $H$ and its resonant cyclotron wave will yield only one Weyl point. In this scenario, the two-bands approximation near the Weyl point and the gap Chern number difference between the two bulks are equivalent. Conversely, if the density varies non-monotonically at the interface while the bulk density remains fixed, the gap Chern number will remain unchanged; however, the number of Weyl points will increase. As outlined in the previous section, the total number of chiral modes at the interface is given by the sum of the Chern numbers contributed by all Weyl points. Consequently, the presence of a localized chiral mode does not necessarily occur near each Weyl point but rather depends on the overall topological properties of these points. Figure \ref{fig4}(e-f) illustrates that although the plasma frequency intersects the R-wave-resonance frequency multiple times, the number of chiral interface modes is always one, which corresponds to the total spectral flow index of all Weyl points. The continuity of $\omega_p$ variation guarantees that adjacent non-trivial Weyl points exhibit opposite spectral flow indices.

In treating the magnetized plasma in this section, the Hamiltonian \( H_0 \) is not regularized. However, the Berry curvature on $B_0$ yields an overall zero result as \( |k| \) tends toward \( +\infty \). Notably, at $B^0$, the curvature at the point \( (x, k_x, k_y) \) is canceled by the curvature at the point \( (x, -k_x, -k_y) \). When incorporating \( H_1 \), the Berry curvature integral over the entire \( B^0 \), still asymptotically approaches zero as \( r \) increases, as shown in Figure \ref{fig:3}(c). 

For other values of the parameter \(k_z\), the plasma frequency \(\omega_p\) corresponding to each of the two degenerate points shifts, yet these points continue to exist in the spectrum. Figure~\ref{fig2_1}.(b) illustrates how the two resonance points (the R-wave in red and the L-wave in blue) move as functions of \(k_z\) and \(\omega_p\). For every fixed \(k_z\), there are two intersections. However, when the overall density—i.e., \(\omega_p\)—becomes sufficiently large, the R-wave resonance point disappears. In such a situation, no chiral interface modes emerge, as shown in Figure~\ref{fig4}.(c-d).

Although the L-wave resonance point still gives rise to a locally nontrivial Berry monopole, producing an integer gap Chern number difference of \(+1\) between the two bulks (which, as shown in the previous section, corresponds to the spectral flow index), no chiral modes can appear if the gap condition between the green and red bands in Figure~\ref{fig2_1}.(a) is not satisfied. 

The R-wave resonance frequency \(\omega_{pc}\) in magnetized warm plasma system is given by
\begin{equation}\label{resonance}
\begin{array}{cc}
\omega_1^2-\omega_p^2-k_z^2 u^2=0,\\
\omega_2^3-\omega_2 (k_z^2+\omega_p^2 )-(\omega_2^2-k_z^2 )=0,\\
\omega_1=\omega_2 := \omega_{pc}> 0,  
\end{array}
\end{equation}
(see Appendix~\ref{two_band}).
The condition for the existence of a chiral mode(TLCWs) in the gap is thus:
\[
(\omega_p^- - \omega_{pc})(\omega_p^+ - \omega_{pc})<0.
\]
For more comprehensive discussions on the topology and symmetry of TLCWs, the reader is referred to \cite{gao2016photonic, fonseca2024first, qin2023topological}.

\section{Conclusions}\label{section4}
This paper establishes a rigorous theoretical framework that connects the spectral flow index—a genuine topological invariant in phase space—to the pseudo-Chern numbers defined in momentum space. The key findings bridge the heuristic application of pseudo-Chern numbers with the rigorous framework of phase-space topology and spectral flow, providing a solid justification for previous results derived through the heuristic bulk-edge correspondence.

We demonstrate that although pseudo-Chern numbers are not true topological invariants, they can be related to the spectral flow index under specific conditions. By defining appropriate connections and curvatures, such as the Berry connection and Berry curvature, we prove that the difference in pseudo-Chern numbers of bulk bands in momentum space equals the spectral flow index. This relationship holds even without regularization in systems with symmetry in momentum space, despite the presence of non-integer bulk Chern numbers.

We also apply this framework to a (2+1)-dimensional finite-temperature magnetized plasma model, emphasizing how symmetry negates regularization requirements while aligning the spectral flow index with differences in gap Chern numbers. It has been demonstrated that the bulk-edge correspondence expressed in our framework is valid, provided that the total integral over the phase-space surface at infinity approaches zero. Moreover, in warm magnetized plasma systems, topological Langmuir-cyclotron waves (TLCW) still exist, and the results are qualitatively similar to those found in studies of magnetized cold plasma. Finally, we present a generalized condition under which TLCW can exist. As a topological edge wave, TLCW can propagate unidirectionally and without reflection or scattering along complex boundaries. Due to its intrinsic topological stability, TLCW could be relatively easy to excite in experimental settings. 

However, real-world plasmas—whether in laboratory or astrophysical environments—are influenced by numerous physical processes not accounted for in this model, such as collisions and non-uniform finite temperature effects. To enable practical applications, these factors must be thoroughly investigated through both experimental approaches and theoretical analyses.

The findings presented in this work provide a robust justification for previous results obtained through heuristic bulk-edge correspondence, bridging the gap between the heuristic use of pseudo-Chern numbers and the rigorous framework of phase-space topology and spectral flow. This research paves the way for a deeper understanding of topological phenomena in continuous media.

\env{acknowledgments} 
% an environment and not a command.
 This work was supported by the National Natural Science Foundation of China [Grant No. 11775220].
\bibliographystyle{apsrev4-2}
%\begin{thebibliography}{00}
\bibliography{reference}

\onecolumngrid
\appendix
\section{Some mathematical tools}\label{mthtool}
\subsection{Weyl quantization \cite{dubin2000mathematical}}
An operator \(a(x,\xi)\in S(\mathbb{R}^n\times \mathbb{R}^n;\mathbb{C})\), called a *symbol* on the phase space \(T^{\star}\mathbb{R}^n = \mathbb{R}^{2n}\), is associated with a pseudo-differential operator (PDO) denoted by \(\widehat{a} = Op_{\epsilon}(a)\). For a function \(\psi \in S(\mathbb{R}^n)\), this operator is defined as:
\begin{equation}
(Op_\epsilon(a)\,\psi)(x)\;=\;\frac{1}{(2\pi\epsilon)^n}\int\,a\!\Bigl(\frac{x+y}{2},\,\xi\Bigr)\,\exp\!\Bigl(\tfrac{i}{\epsilon}\,\xi\cdot(x-y)\Bigr)\,\psi(y)\,dy\,d\xi.
\end{equation}

For a given operator \(\hat{a}\), its symbol can be obtained through the Wigner transform:
\begin{equation}
a(x,\xi)\;=\;\int \,\Bigl\langle\,x+\tfrac{y}{2}\,\Big|\;\hat{a}\bigl(\hat{x},\hat{\xi}\bigr)\;\Big|\,x-\tfrac{y}{2}\Bigr\rangle \,\exp\!\Bigl(-\tfrac{i}{\epsilon}\,\xi\cdot y\Bigr)\,dy,
\end{equation}
where \(\widehat{x} = x\) and \(\widehat{\xi} = -\,i\,\epsilon\,\nabla\).

\subsection{Faure’s index theorem \cite{faure2019manifestation}}

Faure’s Index Theorem relates the spectral flow index to the Chern number (Berry-Chern monopoles) arising from the degenerate points of two bands of a Weyl symbol.

1. Setup: 

   Consider an operator \(\widehat{H}_\mu (x,-i\epsilon \nabla)\) with an Hermite symbol \(H_\mu(x,\xi)\). The symbol is an \(n\times n\) matrix with eigenvalues denoted by
   \[
   \omega_1(\mu,x,\xi)\;\le\;\omega_2(\mu,x,\xi)\;\le\;\cdots\;\le\;\omega_n(\mu,x,\xi).
   \]

2. Spectrum Gap Assumption:

   Suppose there exists an index \(r \in \{1, \ldots, n-1\}\) and a constant \(C>0\) such that for all \((\mu, x, \xi)\) with \(\|(\mu, x, \xi)\|\geq r > 0\), the following strict inequality holds:
   \begin{equation}
       \omega_s(\mu,x,\xi) \;<\; g_1 \;<\; g_2 \;<\; \omega_{s+1}(\mu,x,\xi).
   \end{equation}
   In other words, there is a nonzero gap \(\bigl[g_1, g_2\bigr]\) separating \(\omega_s\) from \(\omega_{s+1}\).

3. Spectral Flow: 

   Vary the parameter \(\mu\) continuously from \(-\infty\) to \(+\infty\). The \emph{spectral flow} \(N\) is defined as the net number of eigenvalues of \(\widehat{H}_\mu (\widehat{x},\widehat{\xi})\) that move from below the band gap \(\bigl[g_1, g_2\bigr]\) to above it (counted with orientation).

4. Chern Number:

   Let \(\omega_1, \omega_2, \ldots, \omega_s\) be the eigenvalues of \(H_\mu(x,\xi)\) lying below the band gap. The corresponding eigenvectors form a vector bundle \(F^-\) over a sphere in phase space that encloses the region where \(\omega_s(\mu,x,\xi)=\omega_{s+1}(\mu,x,\xi)\) (i.e., the gapless points). The integral of the Berry curvature of these eigenvectors over that enclosing sphere yields an integer, called the \emph{Chern number} of the eigenbundle \(F^-\), denoted by \(C\).

5. Faure’s Index Theorem:

   Faure’s result states that
   \[
   N \;=\; C,
   \]
   i.e., the spectral flow index \(N\) equals the Chern number \(C\).

This theorem establishes a topological correspondence: the net count of how many eigenvalues cross the gap (spectral flow) matches the total Berry-Chern charge (Chern number) associated with the degenerate points in the Weyl symbol.

\section{Linearization and normalization of WEMHD}\label{linearization}
Substituting the isothermal relation \(P_e = n\,k_B T_e\) into the WEMHD equations gives:
\begin{subequations}
\begin{eqnarray}
m_e\,n\,\partial_t \boldsymbol{v} &=& -\,n\,e\,\bigl(\boldsymbol{E} + \boldsymbol{v} \times \boldsymbol{B}\bigr)\;-\;k_B\,T_e\,\nabla n, 
\\[6pt]
\partial_t n &=& -\,n\,\nabla \cdot \boldsymbol{v}\;-\;\nabla n \cdot \boldsymbol{v}, 
\\[6pt]
\partial_t \boldsymbol{B} &=& -\,\nabla \times \boldsymbol{E}, 
\\[6pt]
\partial_t \boldsymbol{E} &=& c^2\,\nabla \times \boldsymbol{B} + c^2\,\mu_0\,n\,e\,\boldsymbol{v}.
\end{eqnarray}
\end{subequations}

Next, consider the decomposition \(\boldsymbol{v} = \boldsymbol{v}_0 + \boldsymbol{v}_1\), \(\boldsymbol{E} = \boldsymbol{E}_0(x) + \boldsymbol{E}_1\), \(\boldsymbol{B} = \boldsymbol{B}_0 + \boldsymbol{B}_1\), and \(n = n_0(x) + n_1\), where \(\boldsymbol{v}_0 = 0\), \(n(x)\,e\,\boldsymbol{E}_0(x) + k_B\,T_e\,\nabla n(x) = 0\), and \(\boldsymbol{B}_0 = B_z\,\boldsymbol{e}_z\). Linearizing the WEMHD equations around this equilibrium yields:
\begin{subequations}
\begin{eqnarray}
\partial_{t}\widetilde{\boldsymbol{v}}_{1} &=& -\,\Bigl[\omega_{p}(x)\,\widetilde{\boldsymbol{E}}_{1} + \widetilde{\boldsymbol{v}}_{1}\times\boldsymbol{e}_{z}\Bigr] - \frac{v_{s}}{c}\,\nabla\widetilde{n}_{1} + \frac{1}{2}\,\frac{v_{s}}{c}\,\frac{\nabla n_{0}}{n_{0}}\,\widetilde{n}_{1},
\\[6pt]
\partial_t \tilde{n}_1 &=& -\,\frac{v_s}{c}\,\nabla \cdot \widetilde{\boldsymbol{v}}_1 - \frac{1}{2}\,\frac{v_s}{c}\,\frac{\nabla n_0}{n_0}\cdot \widetilde{\boldsymbol{v}}_1,
\\[6pt]
\partial_{t}\widetilde{\boldsymbol{B}}_{1} &=& -\,\nabla \times \widetilde{\boldsymbol{E}}_{1},
\\[6pt]
\partial_{t}\widetilde{\boldsymbol{E}}_{1} &=& \nabla \times \widetilde{\boldsymbol{B}}_{1} + \omega_{p}(x)\,\widetilde{\boldsymbol{v}}_{1},
\end{eqnarray}
\end{subequations}
where each variable with a tilde denotes a dimensionless, normalized perturbation. Time is normalized by \(1/\Omega\), length by \(c/\Omega\), the electric field \(E_1\) by a reference field \(E\), where  $\Omega = e\,B_z/m_e$ is the electron cyclotron frequency; the magnetic field \(B_1\) by \(E/c\), velocity \(v_1\) by $\sqrt{\epsilon_0 E^2/m_e n_0}$, and density \(n_1\) by $\sqrt{n_0 \epsilon_0 E^2/k_B T_e}$; \(c\) is the speed of light; the function \(\omega_p(x)\) is defined by $\omega_p(x) = \sqrt{n_0 e^2/\epsilon_0 m_e }/(eB_z/m_e )$, and \(v_s = \sqrt{k_B T_e/m_e}\).

%\begin{subequations}\begin{eqnarray}
%&\partial_{t}\boldsymbol{v_{1}}=-\frac{e}{m_{e}}(\boldsymbol{E_{1}}+v_{1} \times \boldsymbol{B_{0}})-%\frac{k_{B}T_{e}}{m_{e}}\nabla\left(\frac{n_{1}}{n_{0}}\right), \\ 
%&\partial_{t}n_{1}=-n_{0}\nabla\cdot\boldsymbol{v_{1}}-\nabla n_{0}\cdot\boldsymbol{v_{1}}, \\
%&\partial_{t}\boldsymbol{B_{1}}=-\nabla\times\boldsymbol{E_{1}}, \\
%&\partial_{t}\boldsymbol{E_{1}}=c^{2}\nabla\times\boldsymbol{B_{1}}+c^{2}\mu_{0}n_{0}e\boldsymbol{v_{1}}.
%\end{eqnarray}\end{subequations}

\section{Closedness of the \((2n-1)\)-Form \(\mathrm{Tr}\bigl(\bigl(U^{-1} dU\bigr)^{2n-1}\bigr)\) for \(U \in U(N)\)}\label{Closed_form} 
From the identity
\[
0 \;=\; d\bigl(U^{-1}U\bigr) \;=\; d\bigl(U^{-1}\bigr)\,U \;+\; U^{-1}\,\bigl(dU\bigr),
\]
and the relation \(U^{-1} = U^{*}\), we deduce
\[
\bigl(U^{-1}\,dU\bigr)^{*}
\;=\;\bigl(dU\bigr)^{*}\,\bigl(U^{-1}\bigr)^{*}
\;=\;\bigl(dU^{-1}\bigr)\,U.
\]
Hence \(U^{-1}(dU)\) is a skew-symmetric one-form, denoted by \(A\). We also have
\begin{equation}
d\bigl(U^{-1}\bigr) \;=\; -\,U^{-1}\,\bigl(dU\bigr)\,U^{-1},
\end{equation}
implying
\begin{equation}
\begin{aligned}
dA \;=\; d\bigl(U^{-1}\,dU\bigr)
&=\;d\bigl(U^{-1}\bigr)\,\wedge\,dU
=\;U^{-1}\,(dU)\,U^{-1}\,\wedge\,dU\\
&=\;-\,\Bigl(U^{-1}\,(dU)\Bigr)^{\wedge2}
=\;-\,A\,\wedge\,A.
\end{aligned}
\end{equation}
Applying the Leibniz rule to \(A^{2n-1}\) gives
\begin{equation}
\begin{aligned}
d\bigl(A^{2n-1}\bigr)
&=\;\sum_{k=1}^{2n-1}(-1)^{\,k-1}\,A^{\,k-1}\,\wedge\,\bigl(dA\bigr)\,\wedge\,A^{\,2n-k-1}
=\;\sum_{k=1}^{2n-1}(-1)^k\,A^{\,2n}\\
&=\;-\,A^{\,2n}.
\end{aligned}
\end{equation}
Since \(A\) is a skew-symmetric form, we can write it locally as
\begin{equation}
A \;=\;\sum_{\mu=1}^{2n}\,A_\mu\,dx_\mu,
\end{equation}
where each coefficient \(A_\mu\) is a skew-symmetric matrix over the complex field. Consequently,
\begin{equation}
\begin{gathered}
A^{2n}
\;=\;\sum_{\mu_1,\mu_2,\dots,\mu_{2n}\in\{1,\dots,2n\}}
A_{\mu_1}\,A_{\mu_2}\,\dots\,A_{\mu_{2n}}\;
dx_{\mu_1}\,\wedge\,dx_{\mu_2}\,\wedge\,\dots\,\wedge\,dx_{\mu_{2n}}\\
=\;\Biggl(\sum_{\mu_1,\dots,\mu_{2n}}
(-1)^{\tau(\mu_1,\mu_2,\dots,\mu_{2n})}\,
A_{\mu_1}\,\dots\,A_{\mu_{2n}}\Biggr)\,
dx_1\,\wedge\,dx_2\,\wedge\,\dots\,\wedge\,dx_{2n}\\
\equiv\;G\,dx_1\,\wedge\,dx_2\,\wedge\,\dots\,\wedge\,dx_{2n}.
\end{gathered}
\end{equation}
Using the cyclic invariance of the trace, we obtain
\begin{equation}
\begin{aligned}
\mathrm{Tr}\,G
&=\;\mathrm{Tr}\!\Biggl(
\sum_{\mu_1,\dots,\mu_{2n}}
(-1)^{\tau(\mu_1,\dots,\mu_{2n})}
\,A_{\mu_1}\,\dots\,A_{\mu_{2n}}
\Biggr)\\
&=\;\tfrac12\sum_{\mu_1,\dots,\mu_{2n}}
\mathrm{Tr}\Bigl(
(-1)^{\tau(\mu_1,\dots,\mu_{2n})}
\,A_{\mu_1}\,\dots\,A_{\mu_{2n}}
\;+\;
(-1)^{\tau(\mu_{2n},\mu_1,\dots,\mu_{2n-1})}
\,A_{\mu_{2n}}\,A_{\mu_1}\,\dots\,A_{\mu_{2n-1}}
\Bigr)
\;=\;0.
\end{aligned}
\end{equation}
In the last step, we use 
\(\,(-1)^{\tau(\mu_1,\dots,\mu_{2n})} = -\,(-1)^{\tau(\mu_{2n},\mu_1,\dots,\mu_{2n-1})}.\)

\section{Berry curvature integral in $B^0$ and gap condition of $H$}\label{Berry}

Writing the symbol \(H(x,k_x,k_y)\) in full matrix form yields:
\begin{equation}
H \;=\;
\begin{pmatrix}
0 & -i & 0 & -\,i\,\omega_p(x) & 0 & 0 & 0 & 0 & 0 & \tfrac{i\,u}{2\,L_n} \,+\,u\,k_x\\
i & 0 & 0 & 0 & -\,i\,\omega_p(x) & 0 & 0 & 0 & 0 & u\,k_y\\
0 & 0 & 0 & 0 & 0 & -\,i\,\omega_p(x) & 0 & 0 & 0 & u\,k_z\\
i\,\omega_p(x) & 0 & 0 & 0 & 0 & 0 & 0 & k_z & -\,k_y & 0\\
0 & i\,\omega_p(x) & 0 & 0 & 0 & 0 & -\,k_z & 0 & k_x & 0\\
0 & 0 & i\,\omega_p(x) & 0 & 0 & 0 & k_y & -\,k_x & 0 & 0\\
0 & 0 & 0 & 0 & -\,k_z & k_y & 0 & 0 & 0 & 0\\
0 & 0 & 0 & k_z & 0 & -\,k_x & 0 & 0 & 0 & 0\\
0 & 0 & 0 & -\,k_y & k_x & 0 & 0 & 0 & 0 & 0\\
-\,\tfrac{i\,u}{2\,L_n} \,+\,u\,k_x & u\,k_y & u\,k_z & 0 & 0 & 0 & 0 & 0 & 0 & 0
\end{pmatrix}.
\end{equation}

The Berry curvature \(F_\omega(x,k_x,k_y)\) of a band \(\omega\), whose corresponding eigenvector is \(\Psi_\omega\), can be written as
\begin{equation}
F_\omega(x,k_x,k_y)
\,=\,
\boldsymbol{e_n}\,\cdot\,\nabla\,\times\!\Bigl(\,\Psi_\omega^*\,\nabla\,\Psi_\omega\Bigr),
\end{equation}
where \(\nabla\) is defined by \(\boldsymbol{e}_n\,\partial_x + \boldsymbol{e}_{k_x}\,\partial_{k_x} + \boldsymbol{e}_{k_y}\,\partial_{k_y}\). The vector \(\boldsymbol{e_n}\) takes the form \(\tfrac{(0,k_x,k_y)}{\|(0,k_x,k_y)\|}\) at \(B^0\), \((1,0,0)\) at \(B^+\), and \((-1,0,0)\) at \(B^-\).

In the bulk regions where \(\,x > d\) or \(x < -\,d\), \(H\) reduces to \(H_0\) (see equation~(\ref{H0handH1})). The operator \(H_0\) satisfies the symmetry
\begin{equation}
H_0(x,-\,k_x,\,-k_y)\,P
\;=\;
P\,H_0(x,\,k_x,\;k_y),
\end{equation}
where \(P\) is a real constant diagonal matrix with \(P^2=1\), specifically
\begin{equation}
P
\,=\,
\begin{pmatrix}
1&0&0&0&0&0&0&0&0&0\\
0&1&0&0&0&0&0&0&0&0\\
0&0&-1&0&0&0&0&0&0&0\\
0&0&0&1&0&0&0&0&0&0\\
0&0&0&0&1&0&0&0&0&0\\
0&0&0&0&0&-1&0&0&0&0\\
0&0&0&0&0&0&1&0&0&0\\
0&0&0&0&0&0&0&1&0&0\\
0&0&0&0&0&0&0&0&-1&0\\
0&0&0&0&0&0&0&0&0&-1
\end{pmatrix}.
\end{equation}
If \(\Psi_\omega(x,k_x,k_y)\) is an eigenvector of \(H_0\) with eigenvalue \(\omega(x,k_x,k_y)\), then
\begin{eqnarray}
\Psi_\omega\bigl(x,-\,k_x,-\,k_y\bigr)
\,=\,e^{\,i\phi}\,P\,\Psi_\omega\bigl(x,k_x,k_y\bigr),\\[4pt]
\omega\bigl(x,-\,k_x,-\,k_y\bigr)
\,=\,\omega\bigl(x,k_x,k_y\bigr),
\end{eqnarray}
where \(\phi\) is a real constant. Hence, on the surface \(B^0\), the Berry curvature satisfies
\begin{equation}
F_\omega\bigl(x,-k_x,-k_y\bigr)
\,=\,
-\,F_\omega\bigl(x,k_x,k_y\bigr).
\end{equation}
On \(B^\pm\), we obtain
\begin{equation}
F_\omega\bigl(x,-k_x,-k_y\bigr)
\,=\,
F_\omega\bigl(x,k_x,k_y\bigr),
\end{equation}
by noting that \(\boldsymbol{e_n}(x,k_x,k_y)=-\,\boldsymbol{e_n}(x,-k_x,-k_y)\) at \(B^0\), whereas \(\boldsymbol{e_n}(x,k_x,k_y)=\boldsymbol{e_n}(x,-k_x,-k_y)\) at \(B^\pm\). Consequently, the total Berry curvature contributed by \(H_0\) on \(B^0\) vanishes.

Next, let us use polar coordinates for \((k_x,k_y)\), so that
\begin{equation}
k_x = k\,\cos\theta, \quad k_y = k\,\sin\theta,
\end{equation}
where \(k = \sqrt{k_x^2 + k_y^2}\). Set \(\epsilon_k = \tfrac{t\,u}{2\,L_n\,k}\) and \(\epsilon = \tfrac{t\,u}{2\,L_n}\). The eigenvalue polynomial of \(H_t = H_0 + t\,H_1\), namely \(\det\bigl(H_t(x,k_x,k_y)-\omega\,I\bigr)\), is an even-degree polynomial in \(\omega\). It can be written as
\(\omega^2\) multiplied by a fourth-degree polynomial in \(\omega^2\). Because of the factor \(\omega^2\), there are two zero eigenvalues, and the even-degree structure implies that \(\pm\omega\) appear as eigenvalue pairs.

Consider the asymptotic behavior for large \(k\). Suppose \(\omega\sim k^s\). If \(s>1\), then \(\det(H_t-\omega I)\sim \omega^{10}\) and there are no roots whose order exceeds 1. If \(s=1\), let \(\omega=k\,\lambda\). Inserting \(\omega=k\,\lambda\) into the characteristic polynomial and factoring out \(k^{10}\,\lambda^4\) gives
\begin{equation}
k^{10}\,\lambda^4\,\Bigl(\lambda^{6} - \bigl(2 + u^2\bigr)\lambda^4 + \bigl(1+2u^2\bigr)\lambda^2 - u^2\Bigr),
\end{equation}
which factors further as
\begin{equation}
k^{10}\,\lambda^4\,(\lambda^2-1)^2(\lambda^2-u^2).
\end{equation}
Hence, for \(\omega\sim k\), there are six non-zero asymptotic roots of \(\det(H_t-\omega I)=0\). For \(0\le s<1\), the polynomial is asymptotically dominated by \(k^6\,u^2\,\omega^2\) and has no roots. Because \(\det(H_t-\omega I)=0\) is a tenth-degree polynomial containing the factor \(\omega^2\), there are two roots if \(s<0\).

Focusing on the leading-order term of the discriminant of \(\det(H_t-\omega I)/\omega^2\) (expanded as a fourth-degree polynomial in \(\omega^2\)) reveals
\begin{equation}
k^{16}\Bigl[u^4\,\bigl(-1 + u^2\bigr)^2\,\bigl(1+4\,k_z^2\,(-1+u^2)^2\bigr)\,\omega_p^4 
+ \mathcal{O}\bigl(\epsilon_k^2\bigr)\Bigr].
\end{equation}
This expression does not vanish, as \(\epsilon_k\ll 1\) for \(k\to +\infty\). It follows that no additional degeneracies arise in \(H_t\) for large \(k\), aside from the two zeros introduced by the factor \(\omega^2\). Label the positive bands of \(H_t\) in ascending order by \(\omega_1,\omega_2,\omega_3,\omega_4\). Only \(\omega_1\) and \(\omega_2\) maintain a nonzero gap, dictated by their asymptotic behavior. This conclusion holds for all \(t\in[0,1]\).

Therefore, for sufficiently large radii \(R\) of \(B^\pm\), the eigenvector bundle of \(H_0\) below the gap between \(\omega_1\) and \(\omega_2\) is homotopically equivalent to the corresponding eigenvector bundle of \(H_0+H_1\) on the surface \(Z_r\).

\section{$H^0$'s spectral flow index calculation in analytic way: Chern number with two-bands approximation near the Weyl point}\label{two_band}

The Weyl point serves as a “Berry curvature monopole” in phase space thereby imparting a non-trivial topology to the eigenbundle. The Chern number of an eigenbundle hosting an isolated Weyl point can be computed analytically via a two-band approximation, as performed in the magnetized cold plasma case \cite{qin2023topological}. We employ this method for \(H_0\) in the following.

In the phase space of \(H_0(x,k_x,k_y)\), the Weyl points lie along the line \(k_x=0\) and \(k_y=0\), as seen in Figure~\ref{fig2_1}. Setting \(k_x=k_y=0\) reduces the eigenvalue formula (23) to
\begin{equation}
\det\bigl(H_0(x,k_x,k_y)-\omega I\bigr)
= \omega^2\,\bigl(\omega^2-\omega_p^2 - k_z^2\,u^2\bigr)\Bigl(\omega^3 - \omega\bigl(k_z^2+\omega_p^2\bigr)
+ \bigl(\omega^2-k_z^2\bigr)\Bigr)\Bigl(\omega^3 - \omega\bigl(k_z^2+\omega_p^2\bigr)
- \bigl(\omega^2-k_z^2\bigr)\Bigr),
\end{equation}
which admits two classes of non-zero roots:

\textbf{Class 1:} \(\omega_1^2 - \omega_p^2 - k_z^2\,u^2 = 0.\) 

The frequency \(\omega_1\) corresponds to the electron Langmuir wave at finite temperature.

\textbf{Class 2:} \(\omega_2^3 - \omega_2\,(k_z^2 + \omega_p^2)\,\pm\,(\omega_2^2 - k_z^2) = 0.\) 

The frequency \(\omega_2\) corresponds to the L wave (use \(+\)) or the R wave (use \(-\)), each exhibiting symmetry for positive and negative branches. 

We focus on the L-wave resonance in the positive-frequency branch. Hence,
\begin{equation}\label{resonance1}
\begin{array}{cc}
\omega_1^2 \;-\;\omega_p^2 \;-\;k_z^2\,u^2=0,\\[4pt]
\omega_2^3 \;-\;\omega_2\,\bigl(k_z^2 + \omega_p^2\bigr) \;+\;\bigl(\omega_2^2 - k_z^2\bigr)=0,\\[4pt]
\omega_1 \;=\;\omega_2 := \omega_{pc} > 0,
\end{array}
\end{equation}
which is the resonance condition that must be met at the plasma interface. Consequently, for a given \(k_z\), the resonance frequency lies between the plasma frequencies of the two bulk regions. Alternatively, one can rearrange the last condition to
\begin{equation}
\bigl(\sqrt{\omega_p^2 + k_z^2\,u^2}\bigr)^3
\;-\;
\sqrt{\omega_p^2 + k_z^2\,u^2}\,\bigl(k_z^2 + \omega_p^2\bigr)
\;+\;\bigl(\omega_p^2 + k_z^2\,u^2 - k_z^2\bigr)
\;=\;0,
\end{equation}
whose solution is \(\,\omega_p=\omega_p(k_z)\).

The eigenvalue \(\,\omega_1=\sqrt{\omega_p^2 + k_z^2\,u^2}\,\) corresponds to the eigenvector
\begin{equation}
|v_1\rangle
=\bigl\{\,0,\;0,\;\omega_1,\;0,\;0,\;i\,\omega_p,\;0,\;0,\;0,\;k_z\,u\bigr\},
\end{equation}
while the eigenvector for \(\omega_2\) is
\begin{equation}
\begin{aligned}
|v_2\rangle
=\{\,i\,(k_z^2 - \omega_{pc}^2),\;k_z^2 - \omega_{pc}^2,\;0,\;\omega_{pc}\,\omega_p,\;-\,i\,\omega_{pc}\,\omega_p,\;0,\;i\,k_z\,\omega_p,\;k_z\,\omega_p,\;0,\;0\}.
\end{aligned}
\end{equation}

We then construct a two-band system:
\begin{equation}
H_2
=\begin{pmatrix}
\langle v_1|H|v_1\rangle & \langle v_1|H|v_2\rangle\\[6pt]
\langle v_2|H|v_1\rangle & \langle v_2|H|v_2\rangle
\end{pmatrix},
\end{equation}
where
\[
H
=\Bigl(H_0 + (\partial_{k_x} H_0)\,\delta k_x + (\partial_{k_y} H_0)\,\delta k_y
+ (\partial_{\omega} H_0)\,\delta\omega_p\Bigr)\bigg|_{k_x=0,\;k_y=0,\;\omega_p=\omega_{pc}},
\]
and \(\delta>0\) is small, with \(\delta k_x^2 + \delta k_y^2 + \delta \omega_p^2 = \delta^2\) defining the sphere \(S_\delta\) where the topology of the eigenbundle is examined.

An explicit form of \(H_2\) is
\begin{equation}\label{eq-H2}
\begin{aligned}
H_2
&=\begin{pmatrix}
\omega_1 \;+\;\tfrac{\omega_p}{\omega_1}\,\delta\omega_p
&
\dfrac{\bigl(i\,\delta k_x + \delta k_y\bigr)\,k_z\,\bigl(\omega_1^2 - u^2\,\omega_2^2\bigr)}
{2\,\omega_1\,\sqrt{\bigl(k_z^2 - \omega_2^2\bigr)^2 + \omega_p^2\,\bigl(k_z^2 + \omega_2^2\bigr)}}
\\[8pt]
\dfrac{\bigl(-\,i\,\delta k_x + \delta k_y\bigr)\,k_z\,\bigl(\omega_1^2 - u^2\,\omega_2^2\bigr)}
{2\,\omega_1\,\sqrt{\bigl(k_z^2 - \omega_2^2\bigr)^2 + \omega_p^2\,\bigl(k_z^2 + \omega_2^2\bigr)}}
&
\omega_2
\;+\;\dfrac{2\,\omega_2\,\omega_p\,\bigl(\omega_2^2 - k_z^2\bigr)}
{\bigl(k_z^2 - \omega_2^2\bigr)^2 + \omega_p^2\,\bigl(k_z^2 + \omega_2^2\bigr)}\,\delta\omega_p
\end{pmatrix}\\
&:=\begin{pmatrix}a & b^*\\[4pt] b & d\end{pmatrix}.
\end{aligned}
\end{equation}
Near the Dirac point \(\omega_1=\omega_2=\omega_{pc}\), the eigensystem of \(H_2\) can be solved explicitly:
\begin{equation}
\begin{gathered}
\lambda_1
=\tfrac12\Bigl(a + d - \sqrt{\,(a-d)^2 + 4\,|b|^2}\Bigr),
\quad
\lambda_2
=\tfrac12\Bigl(a + d + \sqrt{\,(a-d)^2 + 4\,|b|^2}\Bigr),\\
\psi_{1}
=\bigl(\tfrac{a - d - \sqrt{(a-d)^2 + 4\,|b|^2}}{2\,b},\,1\bigr)
\,=:\,\bigl(\Psi_{11},\,1\bigr),
\quad
\psi_{2}
=\bigl(\tfrac{a - d + \sqrt{(a-d)^2 + 4\,|b|^2}}{2\,b},\,1\bigr).
\end{gathered}
\end{equation}
Imposing the resonance condition~\eqref{resonance}, one obtains
\begin{equation}
a - d
=\tfrac{k_z^2\,(1 - u^2)\,\bigl(k_z^2 + \omega_{pc}^2\bigr)\,\omega_p}
{\omega_{pc}\,\bigl(k_z^2 - \omega_{pc}^2\bigr)^2
\;+\;\omega_{pc}\,\omega_p^2\,\bigl(k_z^2 + \omega_{pc}^2\bigr)}
\;\delta\omega
:= \alpha\,\delta\omega,
\end{equation}
where \(\alpha>0\). Hence, the only zero of \(\Psi_{11}\) on the sphere \(S_\delta\) is \(\bigl(\delta\omega,\delta k_x,\delta k_y\bigr)=(\delta,\,0,\,0)\). The Chern number associated with \(\Psi_1\) equals the winding number of \(\Psi_{11}(\delta,\delta k_x,\delta k_y)\) as \((\delta k_x,\delta k_y)\) encircle \((0,0)\) in an anticlockwise manner. Noting that
\begin{equation}
\frac{\psi_{11}}{\lvert\psi_{11}\rvert}
\,=\,-\,\frac{1}{2b}\Big/\left\lvert\frac{1}{2b}\right\rvert
\,=\,e^{-\,\tfrac{i\pi}{2}}\;e^{\,i\theta},
\quad
\text{where}
\quad
e^{\,i\theta}
=\dfrac{\delta k_x \,+\,i\,\delta k_y}{\sqrt{\delta k_x^2 \,+\,\delta k_y^2}},
\end{equation}
the phase of \(\psi_{11}\big/\lvert\psi_{11}\rvert\) increases by \(2\pi\) around that loop, giving a Chern number of \(+1\) for \(\Psi_1\). Therefore, the L-wave Langmuir-cyclotron resonance has Chern number \(+1\). By a similar analysis, the R-wave Langmuir-cyclotron resonance has Chern number \(-1\), although only the R wave satisfies the gap condition.

\section{Numerical method of band structure in Figure 4}

Consider a density (or plasma frequency) profile along the \(x\)-direction as an inhomogeneous parameter. We perform Fourier transforms in the \(y\) and \(z\) directions of equation~(\ref{H_linear}), introducing wave numbers \(k_y\) and \(k_z\). The computational domain extends from \(-L\) to \(L\), where the distribution in \([-L,0]\) is shown on the right-hand side of Figure~4. The parameters in \([0,L]\) exhibit a mirror-symmetric distribution with respect to \([ -L,0]\). Periodic boundary conditions are imposed on both ends of the interval \(\bigl[-L,\,L\bigr]\). We discretize this interval as \(x_i = -L + i\,\delta x\) for \(i=0,1,\dots,2N\), with \(\delta x = L/N\).

Using this discretization, equation~(\ref{H_linear}) takes the following discrete form:
\begin{equation}
\begin{aligned}
&\left\{\begin{array}{l}
\omega\,v_{x, i}
\,=\,-\mathrm{i}\,v_{y, i}
\,+\,\mathrm{i}\,\omega_{\mathrm{p}, \mathrm{i}}\,E_{x, i}
\,+\,\frac{\mathrm{i}\,u}{2\,L_i}\,\frac{n_{x, i+1 / 2}+n_{x, i-1 / 2}}{2}
\,+\,\mathrm{i}\,u\,\frac{n_{\mathrm{x}, i+1 / 2}-n_{\mathrm{x}, i-1 / 2}}{\Delta x},\\[6pt]
\omega\,v_{y, i}
\,=\,\mathrm{i}\,v_{x, i}
\,+\,\mathrm{i}\,\omega_{\mathrm{p}, \mathrm{i}}\,E_{y, i},\\[6pt]
\omega\,v_{z, i}
\,=\,\mathrm{i}\,\omega_{\mathrm{p}, \mathrm{i}}\,E_{z, i},
\end{array}\right. \\[6pt]
&\left\{\begin{array}{l}
\omega\,E_{x, i}
\,=\,-\,\mathrm{i}\,\omega_{\mathrm{p}, \mathrm{i}}\,v_{x, i}
\,+\,k_z\,\frac{B_{y, i+1 / 2}+B_{y, i-1 / 2}}{2}
\,-\,k_y\,\frac{B_{z, i+1 / 2}+B_{z, i-1 / 2}}{2},\\[6pt]
\omega\,E_{y, i}
\,=\,-\,\mathrm{i}\,\omega_{\mathrm{p}, \mathrm{i}}\,v_{y, i}
\,-\,k_z\,\frac{B_{x, i+1 / 2}+B_{x, i-1 / 2}}{2}
\,-\,\mathrm{i}\,\frac{B_{z, i+1 / 2}-B_{z, i-1 / 2}}{\Delta x},\\[6pt]
\omega\,E_{z, i}
\,=\,-\,\mathrm{i}\,\omega_{\mathrm{p}, \mathrm{i}}\,v_{z, i}
\,+\,k_y\,\frac{B_{x, i+1 / 2}+B_{x, i-1 / 2}}{2}
\,+\,\mathrm{i}\,\frac{B_{y, i+1 / 2}-B_{y, i-1 / 2}}{\Delta x},
\end{array}\right. \\[6pt]
&\left\{\begin{array}{l}
\omega\,B_{x, i+1 / 2}
\,=\,-\,k_z\,\frac{E_{y, i+1}+E_{y, i}}{2}
\,+\,k_y\,\frac{E_{z, i+1}+E_{z, i}}{2},\\[6pt]
\omega\,B_{y, i+1 / 2}
\,=\,k_z\,\frac{E_{x, i+1}+E_{x, i}}{2}
\,+\,\mathrm{i}\,\frac{E_{z, i+1}-E_{z, i}}{\Delta x},\\[6pt]
\omega\,B_{z, i+1 / 2}
\,=\,-\,k_y\,\frac{E_{x, i+1}+E_{x, i}}{2}
\,-\,\mathrm{i}\,\frac{E_{y, i+1}-E_{y, i}}{\Delta x},
\end{array}\right. \\[6pt]
&\left\{\begin{array}{l}
\omega\,n_{x, i+1 / 2}
\,=\,-\,\mathrm{i}\,v_{y, i}
\,+\,\mathrm{i}\,\omega_{\mathrm{p}, \mathrm{i}}\,E_{x, i}
\,-\,\frac{\mathrm{i}\,u}{2\,L_i}\,\frac{E_{x, i+1}+E_{x, i-1}}{2}
\,+\,\mathrm{i}\,u\,\frac{E_{x, i+1}-E_{x, i-1}}{\Delta x},\\[6pt]
\omega\,n_{y, i+1 / 2}
\,=\,\mathrm{i}\,v_{x, i}
\,+\,\mathrm{i}\,\omega_{\mathrm{p}, \mathrm{i}}\,E_{y, i},\\[6pt]
\omega\,n_{z, i+1 / 2}
\,=\,\mathrm{i}\,\omega_{\mathrm{p}, \mathrm{i}}\,E_{z, i}.
\end{array}\right.
\end{aligned}
\end{equation}
We adopt the discrete numerical approach by Fu and Qin \cite{fu2021topological} for cold plasma, which preserves Hermiticity in the discrete system.

\end{document}